\documentclass[manuscript=article]{achemso}
\usepackage{graphicx} 
\usepackage{subcaption}

\title{Commment on ``Secondary Nucleation by Interparticle Energies. I. Thermodynamics''}
\author{Rasmus A. X. Persson}
\affiliation{Arbetslag A, Torpskolan, SE-443 41 Lerum, Sweden}
\email{rasmus.persson@lerum.se}

\begin{document}

\begin{abstract}
    We point out an inconsistency in the theoretical treatment of Bosetti and collaborators (Crystal Growth \& Design, 2021, 22.1: 87-97), and then proceed to prove that for monomers interacting by two-body potentials, the presence of an infinite, flat, attractive crystal surface may not induce an increased propensity for cluster formation (that is, secondary nucleation) of the type envisioned by these authors.
\end{abstract}

\section{Original comment}
Bosetti \textit{et al.} \cite{bosetti2021secondary}, interested in explanations for secondary nucleation, consider two thermodynamic states: 
\begin{enumerate}
    \item free solute molecules in the vicinity of a flat surface, and
    \item free solute molecules and a $n$-sized cluster in the vicinity of the same surface. 
\end{enumerate} 
They write  the free energy of formation of this cluster as
$\Delta G_\mathrm{hon}(\Delta \mu, n) + \Delta G_\mathrm{IP}(n, x)$
where the first term is the free energy of formation of the cluster in the homogeneous bulk at a supersaturation $\Delta \mu$ and the second term is a correction accounting for the presence of a flat surface at a distance $x$ from the cluster. There are some theoretical issues with their approach that need to be clarified.

For the correction term, they write,
\begin{equation}
    \Delta G_\mathrm{IP}(n, x) = E_\mathrm{IP}^\mathrm{c}(n, x) - \sum_{i=1}^n E^\mathrm{m}_\mathrm{IP}(x_i) 
\end{equation}
where $E^\mathrm{m}_\mathrm{IP}(x_i)$ is the potential energy of interaction between molecule $i$ and the surface at a distance $x_i$ from it and $E_\mathrm{IP}^\mathrm{c}(n, x)$ is the corresponding quantity for the whole cluster. However, the authors clearly state in their third numbered assumption that
\begin{quote}
the positions of the $n$ molecules in State 1 are \emph{identical} to those of the $n$ molecules forming the $n$-sized cluster in State 2. (p. 89, emphasis added)    
\end{quote}
 Thus, there is no geometric difference between the configuration of monomers interacting with the surface prior to forming the cluster (the terms under the summation sign), and the configuration with the cluster interacting with the surface (the first term). Since the potential energy only depends on the geometry (which is identical), this leads to the conclusion that this correction term vanishes, that is
\begin{equation}
    \Delta G_\mathrm{IP}(n, x) \equiv 0.
\end{equation}

Now, one may do a charitable interpretation and simple disregard their statement that the positions of the (cluster-forming) molecules in State 1 and State 2 are identical. In fact, this is what is implicitly assumed when Bosetti and coworkers \cite{bosetti2021secondary} introduce continuum \emph{approximations} in the accompanying \textit{Supplementary Information} to calculate $E_\mathrm{IP}^\mathrm c$ (these may arguably  be seen as ``non-zero approximations for zero''). However, even more logically, one may assume that the molecules forming the cluster are \emph{closer} together after having formed the cluster than before. The positions are then not identical in States 1 and 2, and we escape the rigorous conclusion that there is no correction for the presence of the surface, even if we are quite unsure of how to calculate the correction in question (we have to take into account the entropy change for the density fluctation).

The free energy of formation of a cluster can in principle be determined by the reversible work (against the potentials of mean force) needed to bring its constituent monomers together from their dispersed state. This is a function of the starting configuration as much as of the final configuration, but no matter the precise definitions of starting and final states, for the presence of a nearby surface to decrease the work in question, it needs to induce attractive force components along the axis separating the nascent cluster and the approaching monomer. This observation leads to a quite general argument which proves that $\Delta G_\mathrm{IP}$ vanishes for all infinitely large flat surfaces as long as the molecular interactions are truncated beyond the two-body terms, and are likely small also for many-body terms since these contributions  decay much quicker with distance than two-body contributions.

\begin{figure}
    
    \begin{subfigure}{0.5\linewidth}
    \centering
    \includegraphics[width=\linewidth]{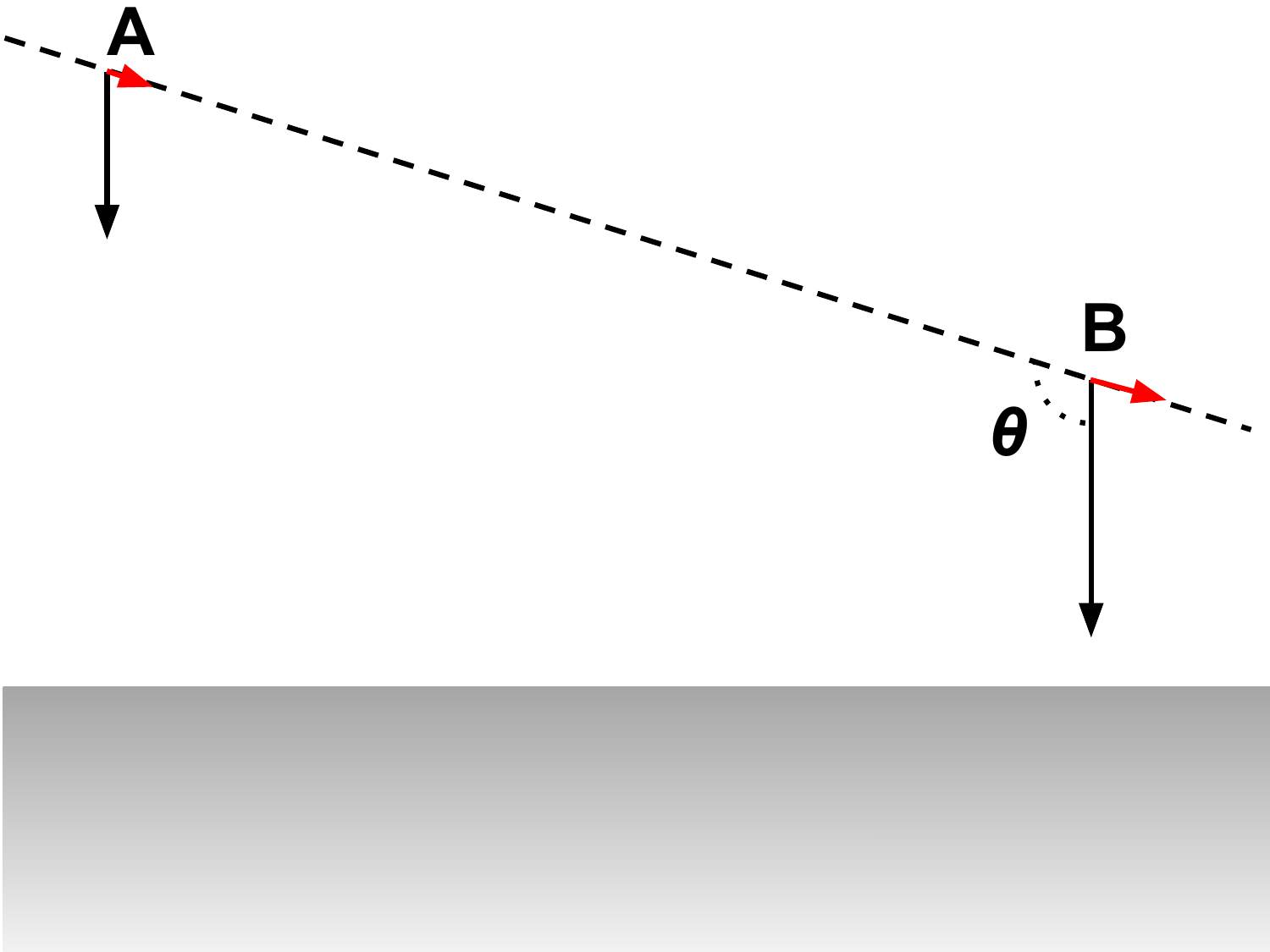}
    \end{subfigure}
    \begin{subfigure}{0.5\linewidth}
    \centering
        \includegraphics[width=\linewidth]{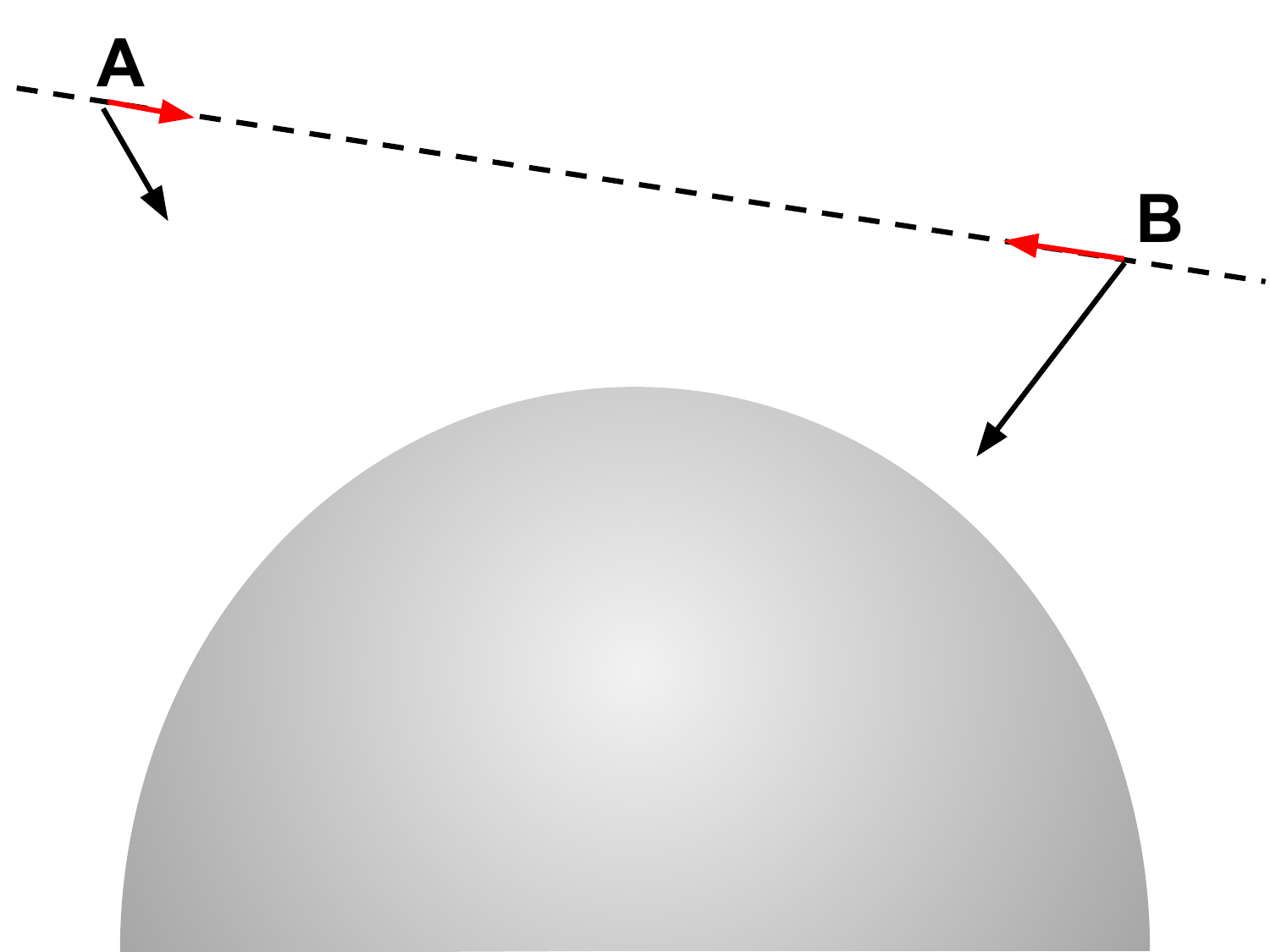}
    \end{subfigure}
    \caption{\textbf{Upper panel:} Forces induced on two monomers by the flat surface (assumed infinitely large) at an angle $\theta$ to the intermolecular axis are indicated by black arrows. The projections of these forces on the intermolecular axis are shown with red arrows. There is never any attraction between the two solutes induced by the surface. \textbf{Bottom panel:} With the curved surface, there may be an attractive force component along the intermonomer axis.}
    \label{fig1}
\end{figure}

If we consider only two-body interactions, symmetry has it that a flat surface may only induce a force orthogonal to itself: \textit{i. e.}, it may either attract or repel along the shortest distance between it and any nearby solute and if the surface is that of a growing crystal, clearly the force is predominantly attractive. Consider now two solute molecules in the vicinity of a macroscopic, flat crystal surface, as illustrated in the top panel of figure~\ref{fig1}. For a secondary nucleation event to occur, these monomers must approach each other to form a nascent cluster. The attractive force induced on the two solutes is, by the symmetry restraint imposed by our assumption of two-body interactions, at an angle $\theta$ with respect to the intermolecular axis of the two solutes. There is a force component of the surface-induced force along this axis that is proportional to the cosine of $\theta$. As long as the closer solute molecule is attracted more strongly to the surface than its more distant counterpart, it is clear by geometry that there is no net attraction along their intermolecular axis for any $\theta$.

In order for there to be a net attraction induced between the two solutes, one must either have a curved surface (as illustrated in the bottom panel of figure~\ref{fig1}) or a spatially oscillatory force of attraction to the crystal surface. The former requires small radii of curvature and the latter situation -- strictly impossible in the gas phase -- is not of any practical significance seeing as the oscillations are (1) quickly damped with distance from the surface and (2) of the length order of the steric size of solvent molecules, thus averaging out to zero over lengths comparable to the intersolute spacing. 

We can very succinctly reinterpret our findings in terms of energies instead of forces, at the expense of a loss of rigor in the argument. In this case, we note that while the free energy of a cluster of molecules will decrease when bringing it into proximity of an attractive surface, so will that of the corresponding free monomers, leading to no change in the actual statistical propensity of cluster formation among them since both ``State 1'' and ``State 2'' are similarly favored by the presence of the surface. (Obviously, judging from purely thermodynamic considerations, the monomers should rather prefer to attach to the existing crystal than to form a new cluster in its vicinity).

Finally, we note that explanations of secondary nucleation that involve mechanical shearing\cite{agrawal2015secondary,zhang2019control,xu2020overview} are impervious to the above arguments.

\bibliography{nuc}

\appendix
\section{Authors' rebuttal}
The above comment was rejected by \emph{Crystal Growth \& Design}. A big part of the reason was the rebuttal by the original authors -- to which I was not given the option to reply -- the main thrust of which is reproduced verbatim below
\begin{quote}
While the geometric contributions are indeed identical in both states, it is essential to include the Hamaker constant to accurately calculate the potential energy. The Hamaker constant is a key parameter to account for the material properties of the interacting bodies and the medium through which they interact. The Hamaker constant – which is different for molecules in solution and for the same molecules in a cluster – must be multiplied to the factor that accounts for the geometric arrangements of molecules – which is the same for molecules and clusters, because we are correctly assuming that the geometrical arrangements are the same. Eq. (2) of the comment, which is based on arguing that ``\emph{Since
the potential energy only depends on the geometry (which is identical)}'', is incorrect because the potential energy depends also on the nature of the phase to which molecules belong, as accounted for by the Hamaker constant; without this, the calculation of potential energy would be incomplete and incorrect.
\end{quote}
I take this rambling argument as proof that my own argument is correct. The potential energy is a function of the coordinates of the molecules. Differences in potential energy between different phases is a function of different molecular arrangements between them. If the molecules do not change their relative positions, then their potential energy remains the same.

\end{document}